\documentstyle[mprocl]{article}

\def\be{\begin{equation}}
\def\ee{\end{equation}}
\def\bea{\begin{eqnarray}}
\def\eea{\end{eqnarray}}
\begin{document}

\title{ Origin  of Structure  in a \\ Supersymmetric 
Quantum Universe}

\author{  P. Vargas 
Moniz}

\address{ 
DAMTP,  
University of Cambridge\\ Silver Street, Cambridge, CB3 9EW, UK\\
{\small e-mail: {\sf prlvm10@amtp.cam.ac.uk; 
paul.vargasmoniz@ukonline.co.uk}}\\
 {\small URL: 
{\sf http://www.damtp.cam.ac.uk/user/prlvm10}}}

\maketitle\abstracts{In this report we 
 advance  the current 
repertoire of quantum cosmological  models to 
incorporate inhomogenous field modes 
in a supersymmetric manner. In particular, we introduce 
perturbations about a
 supersymmetric FRW model.
A quantum state of our model 
has properties typical of the 
 {\em no-boundary} (Hartle--Hawking) proposal. This solution 
 may then lead to a  scale--free spectrum of density perturbations.
%More details can be found in ref. \cite{newnew}.
}

The  main objective of this report  is  to 
examine {\em if} and {\em how} 
the inclusion of supersymmetry 
in a quantum cosmological scenario 
may add to our 
understanding of structure formation in the very early universe. 
Supersymmetry is a
 transformation which relates bosons and fermions and 
 induces  the cancelation 
of divergences that are 
otherwise present in plain quantum gravity.
Thus, its  presence  seems to constitute 
an element of  remarkable  value. 

The fundamental element of our approach
is that 
 N=1 supergravity  constitutes a   ``square-root'' 
\cite{2}
of
 gravity: in finding a physical state $\Psi$,  it is
sufficient to  solve
the Lorentz and supersymmetry constraints; 
 $ \Psi $ will consequently  obey 
 the Hamiltonian constraints\footnote{For a review on  
canonical quantization of supersymmetric 
minisuperspaces  see  
ref. \cite{rev}.}.
In more precise terms, 
 simple {\em first-order}  
differential equations 
have to be solved.
This   contrasts with the situation without supersymmetry:  
 {\it second-order}  Wheeler-DeWitt  equation
 has to be solved,  employing     boundary conditions
\cite{HH83}.

The action for our model is retrieved from the general 
action of the theory of N=1 of supergravity\cite{6} with
 scalar supermultyplets.
 Our {\em background} 
 supersymmetric FRW minisuperspace  is 
described  by a 
tetrad  $e^{AA'}_\mu = e^a_\mu \sigma_a^{AA'}$ with 
$ 
e_{a\mu} = {\rm diag} [N(t), a (t) E_{\hat a i} ]$, 
where   $ \hat a $ and $ i $ run from 1 to 3, 
$ E_{\hat a i} $ is a basis of left-invariant 1-forms on the unit $ S^3 $
and $N(t)$, $a(t)$, $\sigma_a^{AA'} (A = 0,1)$ denote respectively 
the lapse function, scale factor and Infeld--Van der Warden symbols.
The gravitinos must have the form\cite{rev} 
$
\psi^A_{~~i} = e^{AA'}_{~~~~i} \bar\psi_{A'} (t) ~, ~
\bar\psi^{A'}_{~~i} = e^{AA'}_{~~~~i} \psi_A (t) ~, 
$
where  
$\psi_A, \bar\psi_{A'}$ constitute  time-dependent  spinor fields and 
$\psi^A_0 (t), \bar \psi_0^{A'} (t)$ are Lagrange multipliers. 
The ``overline'' denotes Hermitian conjugation. 
A  set of time-dependent  complex scalar fields, 
$\phi, \bar \phi$, and their 
fermionic superpartners, $\chi_A (t), \bar \chi_{A'} (t)$ are also included.

As far as the perturbations about the background 
minisuperspace are concerned, 
we
take the scalar fields as 
\begin{equation}
\Phi (x_i, t)  =  \phi (t) + \Sigma_{nlm} f_{n}^{lm} (t) 
Q^{n}_{lm} (x_i),
\label{eq:essay10a}
\end{equation}
with its Hermitian conjugate 
where 
the coefficients $f_{n}^{lm}$ are functions 
of the time coordinate $t$ and 
$Q^{n}_{lm}$ are standard scalar spherical harmonics\cite{stf1} on 
$S^3$. The fermionic superpartners are expanded as
(see ref. \cite{stf2}): 
\begin{equation}
{\bf X}_A (t, x_i)  =  \chi_A (t) + a^{-3/2} \Sigma_{mpq} 
\beta_m^{pq} \left[ s_{mp} (t) \rho^{nq}_A (x_i) + 
\bar t_{mp} (t) \bar \tau_A^{mq} (x_i) \right],
\label{eq:essay10b} 
\end{equation}
with  $\rho^{mq}_A, 
 \bar \tau^{mq}_{A}$  are 
spinor hyperspherical harmonics on $S^3$.

We obtain   simple 
canonical relations (i.e., 
Dirac brackets) 
\begin{equation}
[\chi_{A}, \bar \chi_{B}]_{D} = -i \epsilon_{AB}~, ~
 [\psi_{A}, \bar \psi_{B}]_{D} = i \epsilon_{AB},   
[a , \pi_{a}]_{D} = 1~, ~ [\phi, \pi_{\phi}]_{D} = 1~,
 ~[\bar \phi, \pi_{\bar \phi}]_{D} = 1,  
\end{equation}
where $\epsilon_{AB}$ is the alternating spinor \cite{rev}. 
All other bracket relations yield zero with the exception of 
\begin{equation}
[ f_n, \pi_{f_{n}} ]_D = \delta_{mn}, ~
[ \bar f_n, \pi_{\bar f_{n}} ]_D = \delta_{mn}, ~
[s_{np}, \bar s_{n'p'}]_{D} = -i\delta_{nn'}\delta_{pp'},
 ~ [t_{np}, \bar t_{n'p'}]_{D} = -i\delta_{nn'} \delta_{pp'}.
\end{equation}
En route to a quantum mechanical description of our model, 
the coordenates of our 
minisuperspace are chosen to be $\left( 
\chi_{A} , \psi_{A} , a , \phi , \bar \phi; f_{n}^{lm}, \bar 
f_{n}^{lm}, s_{np}, t_{np}
\right)$  while $\left( 
\bar \chi_{A} , \bar \psi_{A}, 
\pi_{a}, \right.$ $\left. \pi_{\phi}, \pi_{\bar \phi}; 
\pi_{ f_{n}^{lm}},
\pi_{\bar f_{n}^{lm}}, \bar s_{np}, \bar t_{np}
\right)$   constitute the canonical momentum variables. 

A natural ansatz for the wave function of the universe 
has the form 
\begin{eqnarray}
\Psi & = &   
A + B\psi^C \psi_C + iC \psi^C\chi_C + D \chi^D \chi_D 
+ E \psi^C\psi_C \chi^D \chi_D \nonumber 
\\ & = & A^{(0)}(a,\phi,\bar\phi) 
\Pi_n A^{(n)} (a,\bar \phi, \phi; 
f_n \bar f_n) \Pi_m A^{(m)} 
(a, \phi, \bar \phi, s_{m}, t_{m}) 
\nonumber \\  &  + & 
 B^{(0)}(a,\phi,\bar\phi) 
\Pi_n B^{(n)} (a,\bar \phi, \phi; 
f_n \bar f_n) \Pi_m B^{(m)} 
(a, \phi, \bar \phi, s_{m}, t_{m})    \psi^{C} \psi_{C} 
\nonumber \\ & + &  
C^{(0)}(a,\phi,\bar\phi) 
\Pi_n C^{(n)} (a,\bar \phi, \phi; 
f_n \bar f_n) \Pi_m C^{(m)} 
(a, \phi, \bar \phi, s_{m}, t_{m})    \psi^{C} \chi_{C} \nonumber \\ & + &
D^{(0)}(a,\phi,\bar\phi) 
\Pi_n D^{(n)} (a,\bar \phi, \phi; 
f_n \bar f_n) \Pi_m D^{(m)} 
(a, \phi, \bar \phi, s_{m}, t_{m})   \chi^{C} \chi_{C} \nonumber \\ & + &  
E^{(0)}(a,\phi,\bar\phi) 
\Pi_n E^{(n)} (a,\bar \phi, \phi; 
f_n \bar f_n) \Pi_m E^{(m)} 
(a, \phi, \bar \phi, s_{m}, t_{m})    \psi^{C} \psi_{C} \chi^{D} \chi_{D}, 
\label{eq:2.15}
\end{eqnarray}
where each bosonic coefficient  $A^{(n)}, A^{(m)}$,  ..., 
$E^{(n)}, E^{(m)}$  depends either  on the 
individual perturbation modes $f_{n}$ or $s_{m}, 
t_{m}$. The expression (\ref{eq:2.15}) 
satisfies the Lorentz constraints associated with the 
unperturbed field variables $\psi_A, 
\bar \psi_{A}, \chi_A$ and $\bar \chi_{A}$:
$J_{AB} = \psi_{(A} \bar{\psi}_{B)} - \chi_{(A} \bar{\chi}_{B)}
 =  0$. 
The coefficients 
$s_{m}, t_{m}, \bar s_{m}, \bar t_{m}$,
 will be taken as 
invariant under local 
Lorentz transformation to lowest order in perturbation.
Overall, this approach is fully satisfactory
and indeed  we can extract 
consistent set of solutions.
From the supersymmetry constraints 
$S_A \Psi = 0$ and $\bar S_A \Psi = 0$, 
we then obtain 
(see ref. \cite{newnew} for more details)
the following solutions:
\begin{eqnarray}
E^{(0)} & = & 
\hat{E}_0^{(0)}
\frac{e^{ 3a^2 +  \phi 
(2 \lambda_6 - \Omega_5) - \Omega_5\bar\phi }}{a^{\Omega_6}} 
\label{eq:solt2a} \\~
E^{(n)} & = & 
E_0^{(n)}
e^{-\lambda_7 \bar \phi +  \phi (2 \lambda_8 - \lambda_7)} 
e^{2\lambda_{9} \bar f_n + 2a^2 (n-1) f_n \bar f_n - (\Omega_7 - 
\lambda_{9})  f_n + 
(\Omega_7- \lambda_{9}) \bar f_n},  \label{eq:solt2b} \\
E^{(m)} & = & 
E_0^{(m)}
e^{2\lambda_8 \bar \phi - C_2 \phi \bar \phi - 
\Omega_9  \phi + \Omega_9 \bar \phi } \tilde E,  \label{eq:solt2c}
\end{eqnarray}
where 
$\hat{E}_0^{(0)} = 
E_0^{(0)} e^{-3a^2}$ 
$E_0^{(n)}, E_0^{(m)}$ 
denote integration constants 
and 
 $ \tilde E  \sim s_{mp}$ or 
$t_{mp}$.
The quantities  $\Omega_{1}$, $\Omega_2$, ... represent  
back reactions 
of the scalar and fermionic 
perturbed modes in the homogenous modes and 
are  assumed to be of a very small value. 

Charactheristic features of 
the no-boundary (Hartle-Hawking)  solution 
are  pres\-e\-nt in the  bosonic 
coefficient $E$  
(\ref{eq:solt2a})-(\ref{eq:solt2c}) (see ref. \cite{stf1,stf2,rev}). 
This state  requires $|\Omega_6| \ll 1 $ and the term 
$e^{-n a^2 f_n \bar f_n}, (n \gg 1) $ 
in eq. (\ref{eq:solt2b}) 
to 
dominate over the other remaining exponential terms. 
It seems thus that supersymmetry 
selects  the 
no-boundary (Hartle-Hawking) quantum 
state as mandatory. Since such 
wave function  may lead to a satisfactory 
spectrum of density perturbations 
\cite{stf1}, it thus seems that supersymmetry 
in the  very early universe  intrisically contains the 
relevant seeds for structure formation.

But do the results hereby presented contribute to our understanding 
of the very early universe and if yes, how?
Within our context, 
the answer to those questions is  then a {\em yes} but 
where some  caution is nevertheless required. In conclusion, 
we have shown that by 
employing supersymmetry in a perturbed 
quantum cosmological model  
we can   extract significant information 
about the very early universe evolution. 
Supersymmetric quantum cosmology 
is a forefront line of research  and this 
essay's results point to 
further analysis which surely  will 
have a considerable 
impact on our understanding 
of the very early universe. 

\section*{Acknowledgments}
This work was supported by   a JNICT/PRAXIS--XXI Fellowship BPD/6095/95.  
The author is grateful to  
O. Bertolami, M. Cavaglia, G. Esposito, C. Kiefer, 
S.W. Hawking, 
R. Graham, H. Luckock  and A. Vilenkin for useful conversations 
which further influenced parts of this report and additional 
research. This brief text also represents a short version of an essay 
awarded an Honourable Mention in the 1997 Gravitaty 
Foundation competition.

\end{document}